\documentclass{article}  
\usepackage{lpiabstr_2e} 
\usepackage{times}

\begin{document}
\title{An Effect of Stimulated Radiation Processes on Radio
Emission from Major Planets}
\author{F. V. Prigara}
\affil{Institute of Microelectronics and Informatics, Russian
Academy of Sciences, 21 Universitetskaya, 150007 Yaroslavl,
Russia, (fprigara@imras.yar.ru)}

\runningtitle{RADIO EMISSION: F. V. Prigara}

\titlemake
\begin{abstracttext}

The standard theory of thermal radio emission encounters some
difficulties. The most crucial one is non-possibility to explain
the radio spectrum of Venus in the decimeter range [3]. The radio
spectra of planetary nebulae at high frequencies also are not
comfortably consistent with the standard theory [6]. Here we show
that the account for an induced character of radiation processes
sufficiently improves the predictions of the standard theory.

It was shown recently [5] that thermal radio emission has a
stimulated character. According to this conception thermal radio
emission from non-uniform gas is produced by an ensemble of
individual emitters. Each of these emitters is a molecular
resonator the size of which has an order of magnitude of mean free
path \textit{l} of photons

\begin{equation}
\label{eq1}
l = \frac{{1}}{{n\sigma} }
\end{equation}

\noindent
where \textit{n} is the number density of particles and $\sigma $ is the
absorption cross-section.

The emission of each molecular resonator is coherent, with the wavelength

\begin{equation}
\label{eq2}
\lambda = l,
\end{equation}

\noindent and thermal radio emission of gaseous layer is
incoherent sum of radiation produced by individual emitters.

The condition (2) implies that the radiation with the wavelength
$\lambda $ is produced by the gaseous layer with the definite
number density of particles \textit{n} .

The condition (2) is consistent with the experimental results by
Looney and Brown on the excitation of plasma waves by electron
beam [8,9]. The wavelength of standing wave with the Langmuir
frequency of oscillations depends on the density as predicted by
equation (1). The discrete spectrum of oscillations is produced by
the non-uniformity of plasma and the readjustment of the
wavelength to the length of resonator. From the results of
experiment by Looney and Brown the absorption cross-section for
plasma can be evaluated.

Let us apply the above-formulated condition for emission to the
atmosphere of planet. In this case the number density of molecules
\textit{n }continuously decreases with the increase of height
\textit{h} in accordance with barometrical formula

\begin{equation}
\label{eq3}
n=n_{0}exp(-\frac{{mgh}}{{kT}})
\end{equation}

\noindent where $m$ is the mass of molecule, $g$ is the gravity
acceleration; $k$ is the Boltzmann constant (e.g., [1]). Here the
temperature of atmosphere $T$ is supposed to be constant. In fact
the temperature changes with the increase of height, so the
formula (3) describes the change of molecule concentration in the
limits of layer, for which we can consider the temperature to be
approximately constant (the temperature changes with the increase
of height essentially slower than the concentration of molecules).

 If the emitting particles are the molecules of definite
sort, then their number density is monotonously decreasing with
the increase of height. Therefore, according to the condition for
emission (1), the radio waves with the wavelength $\lambda $ are
emitted by the gaseous layer, located at well-defined height
\textit{h} in atmosphere. Thus, the relation between the
brightness temperature of radio emission and wavelength
\textit{T}($\lambda $) reproduces (partially or in whole) the
temperature section across the atmosphere \textit{T(h)}.

This is the case for the Venus atmosphere. Here the emitting
particles are the molecules of $CO_{2}$. This oxide of carbon is
the main component of Venus atmosphere, its contents being 97\%.
The data concerning the brightness temperature of radio emission
from Venus are summarized in [4].

Assuming $\sigma=10^{-15}cm^{2}$ and using the condition for
emission (1), we can find the number density of emitting molecules
\textit{n} in the gaseous layer, which emits the radio waves with
given wavelength $\lambda $. Then, with the help of barometrical
formula (3) and the data upon the pressure and the temperature in
the lower layer of Venus atmosphere [1], we can establish the
height of emitting layer in the atmosphere. This procedure gives
the temperature section of Venus atmosphere similar to those of
Earth's atmosphere. It is not in contradiction with the data
received by the means of spacecrafts Venera, since on these
apparatus the temperature was measured only in the limits of
troposphere, up to the height 55 km.

Though the direct measurements of a temperature profile at the
heights of 100 to 160 km are absent, the electron density profile
for these heights is available [3]. Using the theory of thermal
ionization [7], one can see that the night profile of electron
density is in agreement with the temperature profile derived in
the present paper.

Quite similarly the brightness temperature of Jovian thermal radio
emission in the range of 0.1 cm to 4 cm as a function of
wavelength [2,4] reproduces the temperature section of Jovian
atmosphere. In this case the temperature structure of atmosphere
also is similar to those of Earth's atmosphere: there are two
minimums of temperature (tropopause and mesopause) and one
intermediate maximum (mesopeak).

The observational data concerning the thermal radio emission of
Mars, Saturn, Uranus and Neptune are not so complete as in the
case of Venus and Jupiter. These data, however, are in agreement
with Earth type temperature structure of atmosphere. In the case
of Mars the measurements of pressure and temperature in the lower
layer of atmosphere are available [1], so one can reproduce using
the radio emission data the temperature section of Mars
atmosphere. Here the emitting particles are the CO$_{2}$ molecules
as in the case of Venus atmosphere.

The intermediate maximum of temperature in the region of mesopeak
perhaps can be explained by absorption and next re-emission of
infrared radiation transferred from the lower layers of atmosphere
[1].

If the planetary atmosphere is sufficiently dense, the own thermal
radio emission from the solid surface of the planet is quenched,
and only the radio emission from the planetary atmosphere is
observed. In accordance with equations (1) and (2), the condition
for the quenching of thermal radio emission with the wavelength
$\lambda$ from a solid surface in a dense atmosphere is as follows

\begin{equation}
\label{eq4}
n_{0}>\frac{{1}}{{\lambda\sigma}}
\end{equation}

\noindent where $n_{0}$ is the number density of molecules nearby
the solid surface. The condition (4) is valid both for Venus and
Mars at radio wavelengths.

Equation (4) determines the quenching of thermal emission from the
solid surface in the gaseous atmosphere in the whole
Rayleigh-Jeans region of spectrum, not only in the radio band.

\begin{center}
REFERENCES
\end{center}

[1] Chamberlain J. W. (1978), Theory of Planetary Atmospheres (New
York: Academic Press). [2] Hubbard W. B. (1984), Planetary
Interiors (London: Van Nostrand Reinhold). [3] Ksanfomality L.V.
(1985), The Planet Venus (Moscow: Nauka). [4] Mayer C. H. (1970),
in Surfaces and Interiors of Planets and Satellites, ed. A.
Dollfus (New York: Academic Press). [5] Prigara F.V. (2003),
\textit{Astron. Nachr., 324, No. S1}, 425. [6] Siodmiak N. and
Tylenda R. (2001), \textit{A\&A, 373}, 1032. [7] Landau L.D. and
Lifshitz E.M. (1976), Statistical Physics (Moscow: Nauka). [8]
Alexeev B.V. (2003), \textit{Usp. Fiz. Nauk, 173}, 145,
\textit{Physics-Uspekhi, Vol.46}. [9] Chen F.F. (1984),
Introduction to Plasma Physics and Controlled Fusion, Vol.1:
Plasma Physics (New York: Plenum Press).

\end{abstracttext}

\end{document}